\documentclass{ws-procs975x65}

\begin{document}

\hskip7.3cm
\noindent{MPP-2007-94, LMU-ASC 49/07}


\vskip0.5cm

\title{STRING THEORY LANDSCAPE AND THE STANDARD MODEL OF PARTICLE PHYSICS\footnote{Contribution to the 11th. Marcel Grossmann Meeting on General Relativity in Berlin, July 2006}}

\author{DIETER L\"UST}

\address{Arnold-Sommerfeld-Center for Theoretical Physics\\
Department f\"ur Physik, Ludwig-Maximilians-Universit\"at M\"unchen\\
Theresienstra\ss e 37, 80333 M\"unchen, Germany
\\
and\\
Max-Planck-Institut f\"ur Physik\\
F\"ohringer Ring 6, 80805 M\"unchen, Germany
\email{luest@theorie.physik.uni-muenchen.de,luest@mppmu.mpg.de}}


\begin{abstract}
In this paper we describe ideas about the string landscape, and how to relate it to the physics of
the Standard Model of particle physics. First, we give a short status report about heterotic
string compactifications. Then we focus on the statistics of D-brane models, on
the problem of moduli stabilization, and finally on some attempts to derive a probability
wave function in moduli space, which goes beyond the purely  statistical count of string vacua.
\end{abstract}

\bodymatter

\section{Introduction}\label{intro}

String theory is a subject with beautiful  connections to mathematics,
in particular to differential and algebraic geometry. 
Moreover, it is well known that string theory provides a consistent
formulation of quantum gravity, and one of its nicest achievements
is the microscopic count of black hole entropies in string theory that precisely reproduces
the thermodynamic area law of Bekenstein and Hawking including corrections from higher
derivative gravity couplings (see \cite{Mohaupt:2007mb} for a recent review).
However not only gravity is an automatic outcome of string interactions,
but also gauge interactions follow from the string world sheet formulation.  
In addition,  there exist a deep relation between gauge and gravitational
interactions in string theory, which manifests itself in the form of the AdS/CFT correspondence \cite{Maldacena:1997re,Witten:1998qj,Gubser:1998bc},
namely the duality between quantum gravity
in 5-dimensional anti-De Sitter space ($AdS_5$) and conformal gauge theories in four-dimensions
on the boundary of $AdS_5$ (the holographic principle).

However, still one of the most important issues is
how to relate string theory to the observables in the low energy physics world.
Specifically, since we know that string theory provides an unification of all
interactions and particles, we want to know how the Standard Model (SM) of elementary
particles with gauge group $G=SU(3)\times SU(2)\times U(1)_Y$ and with 3 generations
of quark and leptons follows from string theory. Next, string theory should provide a framework
for computing all couplings of the SM. 
Space-time supersymmetry is automatically built into string theory; hence we have to
find out how supersymmetry gets broken in string theory, and what are the consequences
of a possible low energy supersymmetry breakdown for the physics at the LHC. In fact, it is
one of the most interesting problems in string phenomenology to provide a translation between
the low energy effective string action and the chain of data, which is expected from the LHC.
Finally, it would be of course most desirable for string theory to provide new testable
experimental signatures, like extra dimensions or the discovery of microscopic black holes
and their Hawking radiation, supporting in this way string theory (or providing arguments
for abandoning it).

One possible approach to string theory is the {\sl top-down} approach, which starts from the unification
of gravity and gauge interactions at very high energies, and then tries to deduce all
low energy observables from investigating the mathematical structures of the theory.
Although we do not yet know the typical string scale $M_{\rm string}$, where the
unification of gravity and gauge interaction takes places, one often assumes, at least
from a conservative point of view, that this happens at the Planck scale of about $M_{\rm Planck}
\simeq 10^{19}~{\rm GeV}$
(there are also alternatives models with much lower string scale \cite{Arkani-Hamed:1998rs}). However no direct experiments
will guide us through the physics at such high energies, a fact which makes the top-down approach
very troublesome. Another problem for the top-down approach is the fast proliferation of
string solutions, the so-called string landscape problem, which we will describe in the
following. The string landscape is defined to be the space of all possible solutions
of the string equations of motion. In ten space-time dimensions, there exist just five different
formulations of string theory (two heterotic strings, type I type IIA and type IIB superstrings).
Exploring several kind of duality symmetries, it is conjectured that all these string theories
can be unified into M-theory, where also 11-dimensional supergravity is included.
However the number of lower-dimensional string solutions, i.e. lower dimensional string
groundstates, which are obtained after compactification, is enormous.
Already in 1984, within  the covariant lattice construction \cite{Lerche:1986cx}, the number of
possible four-dimensional string ground states was estimated to be of order $10^{1500}$.
More recently, the number  of discrete flux vacua on a generic Calabi-Yau manifold was shown
to be of order $10^{500}$ \cite{Bousso:2000xa,Douglas:2003um}.
Taken seriously, this vast landscape of distinct string vacua
really implies a big question
mark concerning the predictivity of string theory, since each point in the landscape essentially
corresponds to a different universe with different particle physics and cosmological properties.
To deal with such a huge number of possibilities, certain strategies are
required in order to proceed within the top-down approach.
One possible and legitimate approach is given by the investigation of the statistical properties
of the string landscape. I.e. one has to determine by statistical methods what is the fraction
of string vacua with good phenomenological properties. Possible statistical correlations
resp. anti-correlations would be especially worth to be discovered, like e.g. between the
number of families and the rank of the low-energy gauge groups, because they could
provide a step towards verifying or resp. falsifying string theory.
In addition, it was argued that the landscape could be the clue for statistically explaining the
hierarchy problem in physics, and especially the smallness of the observed cosmological constant.
Namely, the tiny observed value of $\Lambda/M_{\rm Planck}^4\simeq 10^{-120}$
could be statistically explained \cite{Weinberg:1987dv} if the number of vacua is of order
of $10^{120}$, which indeed seems
to be the case in string theory. Eventually, the statistical approach is likely to be merged with the anthropic principle \cite{Susskind:2003kw} (see also \cite{Schellekens:2006xz}).
Concerning the evolution of the universe (see e.g. \cite{Linde:2007fr}), the anthropic principle essentially requires
a multiverse with a huge number of bubbles, with each being  filled by one of the vacua of the landscape. The population of all possible bubbles in the universe
is possible in the context of eternal inflation, where transitions between different bubbles
due to quantum tunneling processes are going to happen. 

Another useful insight into the landscape discussion might be provided by the quantum
mechanical wave function (Hartle-Hawking wave function \cite{Hartle:1983ai})
of the universe, which describes a probability measure, namely the likelihood for each point in
the landscape to be populated. In string theory, it is natural to assume that this probability
measure is linked to an entropy functional in the landscape, i.e. the number of microscopic
string excitations that correspond to each ground state. Clearly, this point of view
is reminiscent to the discussion of black hole entropies, where the macroscopic black
hole entropy can be explained by counting the corresponding microscopic string degrees
of freedom. We will come to an explicit proposal how to relate black hole entropies to
probability functions in the flux landscape at the end of this paper.

Complementary to the top-down efforts, the {\sl bottom-up} approach is very important
 for connecting string theory with the real world. Here one tries to build consistent 
string models which contain as many SM features as possible. 
First one tries to build string models that contain as massless
states the particles of the SM, gauge bosons and three families of
quarks and leptons. Next, one has to derive the low-energy effectice
action of the massless fields, in order to compute their couplings, like gauge couplings
and Yukawa couplings, which eventually can be compared with the experimentally known values.
Here another problem has to be solved, namely the problem of
moduli stabilization. String compactifications generically contain several massless
moduli fields with flat potential, which correspond to geometrical or other parameters
of the internal space. These have to be fixed, since the low-energy couplings of the
massless fields are functions of the moduli. In order to make predictions
one has to know the values of the moduli. In addition, massless moduli would
overclose the universe and also cause un-observed new forces \cite{deCarlos:1993jw}.

One additional guiding
principle for string model building is the fact that, although the energy scale of the SM is
around $100~{\rm GeV}$, the SM is not a complete model but certainly requires
new physics at mass scales much higher than $100~{\rm GeV}$.
In fact it is encouraging that several puzzles of the SM, most notably the hierarchy problem
together with the problem of the missing dark energy, can be explained in one stroke,
namely by low-energy space-time supersymmetry around the corners at about $1-10~{\rm TeV}$.
This fact gave rise to the formulation of the minimal supersymmetric standard model
(MSSM) or some variants thereof. Since supersymmetry is built into string theory from the very beginning, the discovery of low energy supersymmetry would be also an additional
argument in favor of string theory.
The smallness of neutrino masses can be nicely explained by the see-saw mechanism,
pointing to another new energy scale between $10^{10}-10^{12}~{\rm GeV}$.
Finally, the unification of gauge coupling constants  of the SM happens at the GUT scale
of $10^{16}~{\rm GeV}$,
and gauge coupling apparently works best within the MSSM.
Therefore neutrino masses and gauge coupling unification should be another
ingredient for string model building. 

Another window into new physics beyond the SM comes from astrophysics and cosmology.
Beautiful experiments, most notably COBE and WMAP, provided a precise
image of cosmic microwave background radiation including its small
density variations. In this way the  inflationary scenario of the early universe is now
established as the standard model for cosmology. In addition, we know from
the astrophysical measurements that our universe is spatially flat. Its energy density
is dominated to about $74\%$ by a dark energy component, which behaves very similarly
to a positive cosmological constant. The explanation of this mysterious dark energy is one
of the biggest challenges for astroparticle physics, and hence also for string theory.
The remaining $26\%$ of the energy density is split into so far directly undiscovered dark matter particles (WIMPS), which account for $22\%$  of the total
energy density, and into a left-over $4\%$ component of visible SM
matter fields. Many properties of the dark matter fields are still unknown, although one very promising
candidate for dark matter is the lightest supersymmetric particle (LSP) in the MSSM.
So, strings also should be helpful to identify the nature of dark matter.

In summary, successful  string model building must  take into account all these phenomenological
boundary conditions coming from the SM, from particle physics beyond the SM and also from cosmology.
The top-down constructions which  start from the geometry of the
compactification space must go hand in hand with the bottom-up approach, where
one is guided by  the phenomenological data. In this way, a (not necessarily one-to-one) map
between geometrical and topological properties of the compactifications spaces and
the particle physics observables will be provided. This dictionary between geometry/topology
and particle physics/cosmology is one of the most interesting aspects of string theory, and
will be demonstrated in this paper  by several examples.

The outline of the paper is the following:
in the next section we discuss heterotic string compactifications, and especially new
vector bundle constructions which are helpful in order to derive the spectrum
and the particle content of the MSSM.
In section three  type IIA intersecting D-brane models will be reviewed. We will
address the statistical properties of these models, i.e. we discuss the
question, how many consistent D6-brane embeddings on a given closed string
background  exist in total, and what is the fraction of them that come
close the MSSM. 
Next we will consider the problem of moduli stabilization and also of supersymmetry breaking
by adding background fluxes and taking into account effects from non-perturbative
string instantons. Finally, in section four, we will discuss some aspects about
the relation between flux vacua and black hole entropies, and the related problem
how to formulate for flux compactificatons a wave function in moduli space.

\section{Heterotic string compactifications}\label{het}

In ten dimensions there exist two heterotic strings models with ${\cal N}=1$ supersymmetry
(16 supercharges) and  gauge groups, $G_{10}=E_8\times E_8,SO(32)$.
Compactifications of the heterotic string that preserves ${\cal N}=1$ supersymmetry
in four dimensions (4 supercharges) are built following essentially two steps:

\begin{itemize}

\item  A suitable choice of an internal, six dimensional manifold ${\cal M}_6$, which can be 
a Calabi-Yau (CY) manifold, a supersymmetric orbifold, or, in more complicated cases with background
fluxes, a space with $SU(3)$ group structure. We will focus on Calabi-Yau compactifications.

\item 
A suitable choice of a stable vector bundle $V$ with structure group $H$ over ${\cal M}_6$.
In order to preserve space-time supersymmetry, $V$  must be holomorphic, its fields
 strength has to satisfy
 \begin{equation}
 F_{ab}=F_{\bar a\bar b}=g^{a\bar b}F_{a\bar b}=0\, ,\label{holv}
 \end{equation}
 where $g^{a\bar b}$ is the inverse metric of the complex space ${\cal M}_6$.
 Furthermore one has to demand the absence of tadpoles, i.e. $dH_3=0$, with $H_3$
 being the NS 3-form field strength. This implies the vanishing of the second Chern class
 of $V$:
  \begin{equation}
 c_2(V)=c_2(T)\, ,\label{c2v}
 \end{equation}
 where $c_2(T)$ is the second Chern class of the tangent bundle of the CY-manifold.
 The observed gauge group $G_4$ in four dimensions is given by the commutant of $H$ in $G_{10}$:
  \begin{equation}
 G_{10}\supset H\times G_4\, .\label{hetgauge}
 \end{equation}
 Finally, the number of chiral, massless matter fields (in a given representation of $G_4$ -- see later
 for more details)
 is determined by the third Chern number of $V$:
  \begin{equation}
 N_F={1\over 2}c_3(V)\, .\label{thirdc}
 \end{equation}
 \end{itemize}
 
 Now there exists essentially two classes of heterotic bundle constructions in the
 literature:

 \vskip0.3cm
 \noindent
 {\sl (i) $H$ being a simple Lie-group:}
 
 \vskip0.3cm
 \noindent
These bundles were explicitly constructed by a number of groups,
\cite{Friedman:1997ih,Ovrut:2002jk,Braun:2004xv,Braun:2005ux,Ovrut:2006yr,Andreas:1998ei,Curio:1998vu,Andreas:1999ty,Curio:2004pf,Andreas:2006dm,Andreas:2007ei}
where
most of the explicit constructions  use as a CY base space an elliptically fibred 
Calabi-Yau manifold.
Since $H$ is simple, the four-dimensional gauge group is simple, too. Hence this construction
is very suitable for construction GUT models in four dimensions. E.g.:
\begin{eqnarray}
&~&H=SU(3) \quad\longrightarrow \quad G_4=E_6\, ,\nonumber\\
&~&H=SU(4) \quad\longrightarrow \quad G_4=SO(10)\, ,\nonumber\\
&~&H=SU(5) \quad\longrightarrow \quad G_4=SU(5)\, .
\end{eqnarray}
It is a generic feature of these vector bundles, that the four-dimensional spectrum does not
contain Higgs fields in the adjoint representation of $G_4$. Therefore, in order to further break these GUT
groups to the gauge group of the SM, one needs additional discrete Wilson lines.
In fact, following this method, three generation models without exotic particles can be constructed.

 \vskip0.3cm
 \noindent
 {\sl (ii) $H$ being a non-simple Lie-group:}
 
 \vskip0.3cm
 \noindent
 Here, the structure group $H$ contains at least one Abelian $U(1)$ factor, i.e. it is of the form: $H=H'\times U(1)$
  \cite{Blumenhagen:2005pm,Blumenhagen:2006ux,Blumenhagen:2006wj,Weigand:2006yj}. It follows that also $G_4$ contains the same $U(1)$ gauge factor:
 \begin{equation}
 G_4=G_4'\times U(1)\, .
 \end{equation}
 Of course, this feature is required, if one likes to construct heterotic CY compacatifications,
 which directly lead to the  SM gauge group, or other models with $U(1)$ factors,
 like flipped $SU(5)$ GUT's with gauge group $G_4=SU(5)\times U(1)$.
 Explicit models can be again constructed on elliptically fibred CY-spaces.
 A detailed analysis of the topology of these vector bundles reveals the following interesting
 relation between the Euler numbers $\chi$ of $V$ and the number of SM particle
 representations in four dimensions \cite{Blumenhagen:2006wj}:

\begin{table}[htb]
\renewcommand{\arraystretch}{1.5}
\begin{center}
\begin{tabular}{|c||c|c|}
\hline
\hline
$SU(3)\times SU(2)\times U(1)_Y \times E_7$ & bundle & SM part. \\
\hline \hline
$ ({\bf 3},{\bf 2}, {\bf 1})_{1\over 3}$ & $\chi(V)=g$ & $q_L$ \\
$ ({\bf 3},{\bf 2}, {\bf 1})_{-{5\over 3}}$ & $\chi(L^{-1})=0$ & $-$ \\
\hline
$ ({\bf \overline 3},{\bf 1}, {\bf 1})_{2\over 3}$ & $\chi(\bigwedge^2 V)=g$ & $d^c_R$ \\
$ ({\bf \overline 3},{\bf 1}, {\bf 1})_{-{4\over 3}}$ & $\chi(V\otimes L^{-1})=g$ & $u^c_R$ \\
\hline
$ ({\bf 1},{\bf 2}, {\bf 1})_{-1}$ & $\chi(\bigwedge^2 V\otimes L^{-1})=g$ & $l_L$ \\
$ ({\bf 1},{\bf 1}, {\bf 1})_{2}$ & $\chi(V\otimes L)+\chi(L^{-2}) =g$ & $e^c_R$ \\
$({\bf 1},{\bf 1},{\bf 56})_{1}$ & $\chi(L^{-1})=0$ & $-$ \\
\hline
\end{tabular}
\caption{\small Massless spectrum of $H=SU(3)\times SU(2)\times U(1)_Y$ models 
      with $g={1\over 2}\int_{\cal M} c_3(V)$.
}
\label{signsbc}
\end{center}
\end{table}

\noindent
Various concrete models leading to $\chi(V)=g=3$ were constructed in \cite{Blumenhagen:2006wj}.
Note that in these heterotic compactifications with Abelian vector bundles
there can be more than one $U(1)$ gauge group factors that become massive
due to the coupling to pseudo scalar moduli  fields. This is analog to type II orientifolds,
which we will discuss in the next section. In fact, heterotic compactifications
with Abelian vector bundle are expected to be the string dual of type I compactifications
with magnetized D9-branes.

Let us close the short overview over new heterotic compactifications with a few more remarks.

\begin{itemize}

\item Gauge coupling unification of the SM gauge coupling constants typically occurs in perturbative
heterotic compactifications at a scale of order $M_{GUT}\simeq g_{\rm string}M_{\rm string}\simeq
10^{17}~{\rm GeV}$. This is about one order of magnitude too high compared with
the phenomenologically preferred GUT-scale. However this constraint can be relaxed
by taking into account moduli dependent, perturbative gauge threshold corrections \cite{Dixon:1990pc,Ibanez:1991zv,Ibanez:1992hc,Nilles:1995kb}, or 
including non-perturbative corrections in heterotic M-theory \cite{Witten:1996mz,Nilles:1997vk}.
Other alternatives like models with non-standard hypercharge embeddings are summarized
in \cite{Dienes:1996du}.

\item
Heterotic moduli stabilization, including all vector bundle moduli, is still a quite difficult problem.
For some attempts of heterotic flux compactifications see \cite{Strominger:1986uh,LopesCardoso:2002hd,Becker:2003yv,Becker:2003sh,Curio:2005ew}, and
for the the issue of stabilizing  moduli in heterotic M-theory consult \cite{Braun:2006th}.

\item Recently, phenomenologically very attractive heterotic models have been constructed on
orbifold spaces \cite{Raby:2005vc,Buchmuller:2005jr,Buchmuller:2006ik} and also
using free fermions \cite{Faraggi:2003yd,Faraggi:2004rq,Faraggi:2006bc}, and aspects of the corresponding heterotic landscape were discussed in \cite{Dienes:2006ut,Lebedev:2006tr,Dienes:2007ms}.

\end{itemize}

\section{Intersecting D-brane models}\label{intersec}

Now let us turn to type II orientifold compactifications to four-dimensions on six-dimensional
manifolds ${\cal M}_6$, which we first discussed in       
\cite{Bachas:1995ik,Blumenhagen:2000wh, Angelantonj:2000hi,Aldazabal:2000dg,Cvetic:2001tj}.
In order to incorporate non-Abelian gauge interactions and to obtain
massless fermions in non-trivial gauge representations, one has to introduce
D-branes in type II superstrings. Specifically there exist three classes of
four-dimensional models:

 \vskip0.3cm
 \noindent
 {\sl (i) Type I compactifications with D9/D5 branes:}
 
 \vskip0.3cm
 \noindent
This class of IIB models contain different stacks of D9-branes, which wrap the entire space ${\cal M}_6$, and which
also possess open string, magnetic,
Abelian gauge fields $F_{ab}$ on their world volumes (magnetized branes).
In other words, $F_{ab}$ corresponds to open string vector bundles, and this class of models
is string dual to heterotic string compactifications. For reasoning of Ramond tadpole cancellation,
one also needs an orientifold 9-plane (O9-plane). In addition one can also include D5-branes and
corresponding O5-planes. In the heterotic dual description the
D5/O5 open strings correspond to the non-perturbative sector of the theory. Since the open string gauge fields $F_{ab}$ induce mixed boundary conditions on the D-branes, the internal compact space can be regarded as a non-commutative
space.

\vskip0.3cm
 \noindent
 {\sl (ii) Type IIB compactifications with D7/D3 branes:}
 
 \vskip0.3cm
 \noindent
Here we are dealing with different stacks of D7-branes, which wrap different internal 4-cycles,
which intersect each other. The D7-branes can also carry non-vanishing open string
gauge flux $F_{ab}$. In addition, one can also allow for D3-branes, which are located at different
point of ${\cal M}_6$. In order to cancel all Ramond tadpoles one needs in general O3- and O7-planes.

\vskip0.3cm
 \noindent
 {\sl (iii) Type IIA compactifications with D6 branes:}
 
 \vskip0.3cm
 \noindent
 This class of models contains intersecting D6-branes, which are wrapped around 3-cycles
 of ${\cal M}_6$. Now, orientifold O6-planes are needed for Ramond tadpole 
 cancellation.\footnote{Also coisotropic D8-branes \cite{Font:2006na} can be in principle included, which we
 however leave out from the discussion.}

 \vskip0.3cm
 \noindent
D-brane models of these three different classes generically can be mapped onto each other by T-duality,
resp. IIA/IIB mirror symmetry including open strings and D-branes, and therefore are
essentially on equal footing. Hence, in the following
we will concentrate on IIA intersecting D6-brane models (class (iii)).
Since we want to engineer the SM, it turns out that one needs at least four stacks of D6-branes
\cite{Ibanez:2001nd,Blumenhagen:2001te}:

\vskip0.5cm
\begin{center}
\begin{tabular}{|c|c|c|c|}\hline
{Stack a:}& $N_a=3$ & $SU(3)_a\times U(1)_a$
& QCD branes
\\ \hline
{Stack b:} & $N_b=2$ & $SU(2)_b\times U(1)_b$&
weak branes
\\ \hline
{Stack c:}& $N_c=1$&  $U(1)_c$&
right brane
\\ \hline
{Stack d:} & $N_d=1$ & $U(1)_d$&
leptonic brane
\\ \hline
\end{tabular}
\end{center}
\vskip0.5cm

\noindent
Here, $N_a$ denotes the number of D6-branes in each stack.
The intersection pattern of the four stacks of D6-branes can be
depicted in the next figure:
\vskip0.5cm
\begin{center}
\begin{picture}(0,0)(0,0)
\put(-100,-100){
   \includegraphics[scale=.30]{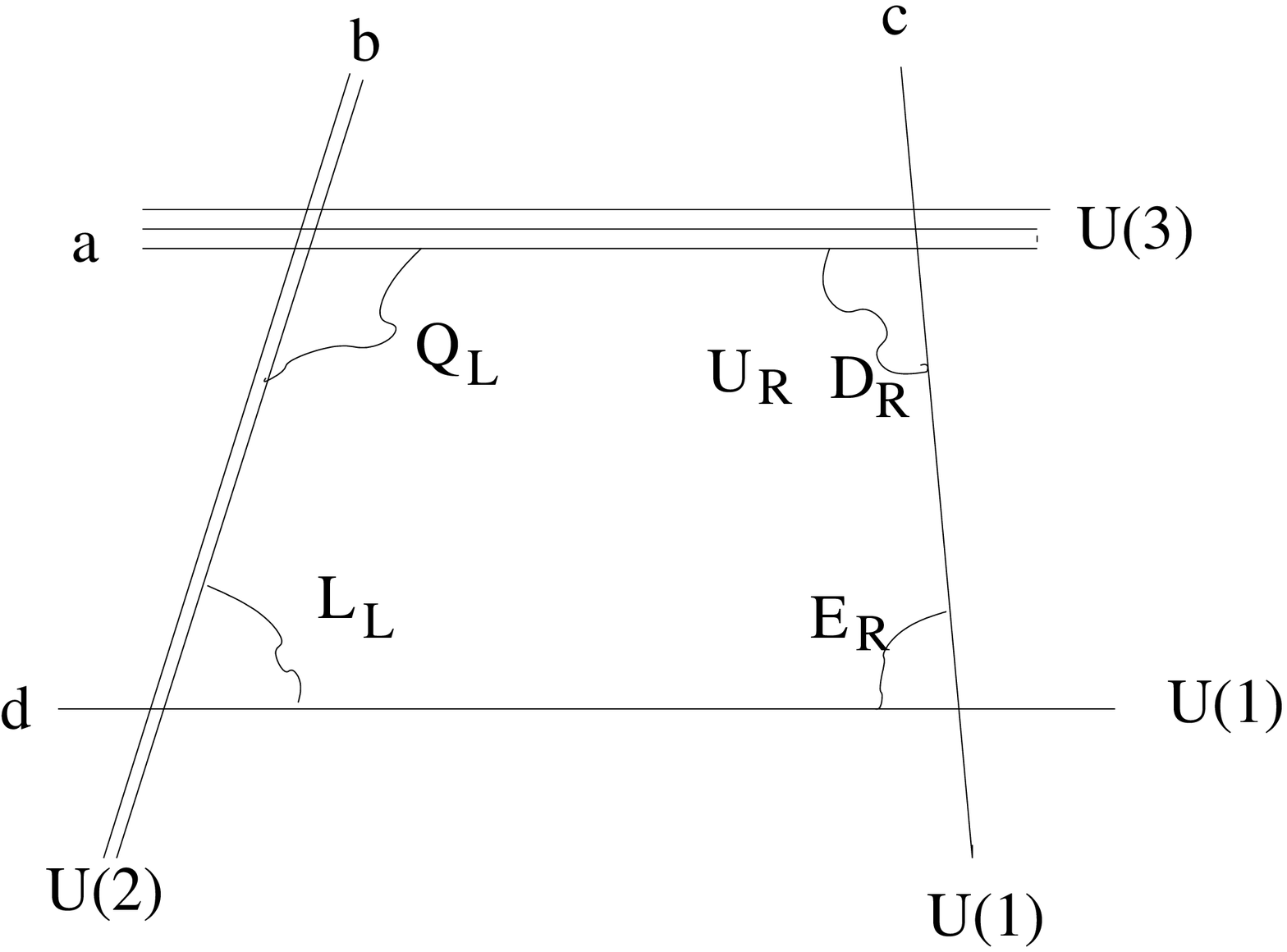}
   }
\end{picture}
\end{center}

\vskip3.3cm
\noindent
This local stack of SM intersecting D6-branes has to be embedded into the compact space
${\cal M}_6$. For reasons of Ramond tadpole conditions one usually needs
more D-branes than this SM brane configuration. These additional D-branes build the
socalled hidden gauge sector. 
This scenario can be depicted in the following picture of the internal space:

\vskip2.5cm
\begin{center}
\begin{picture}(0,0)(0,0)
\put(-100,-100){
   \includegraphics[scale=.35]{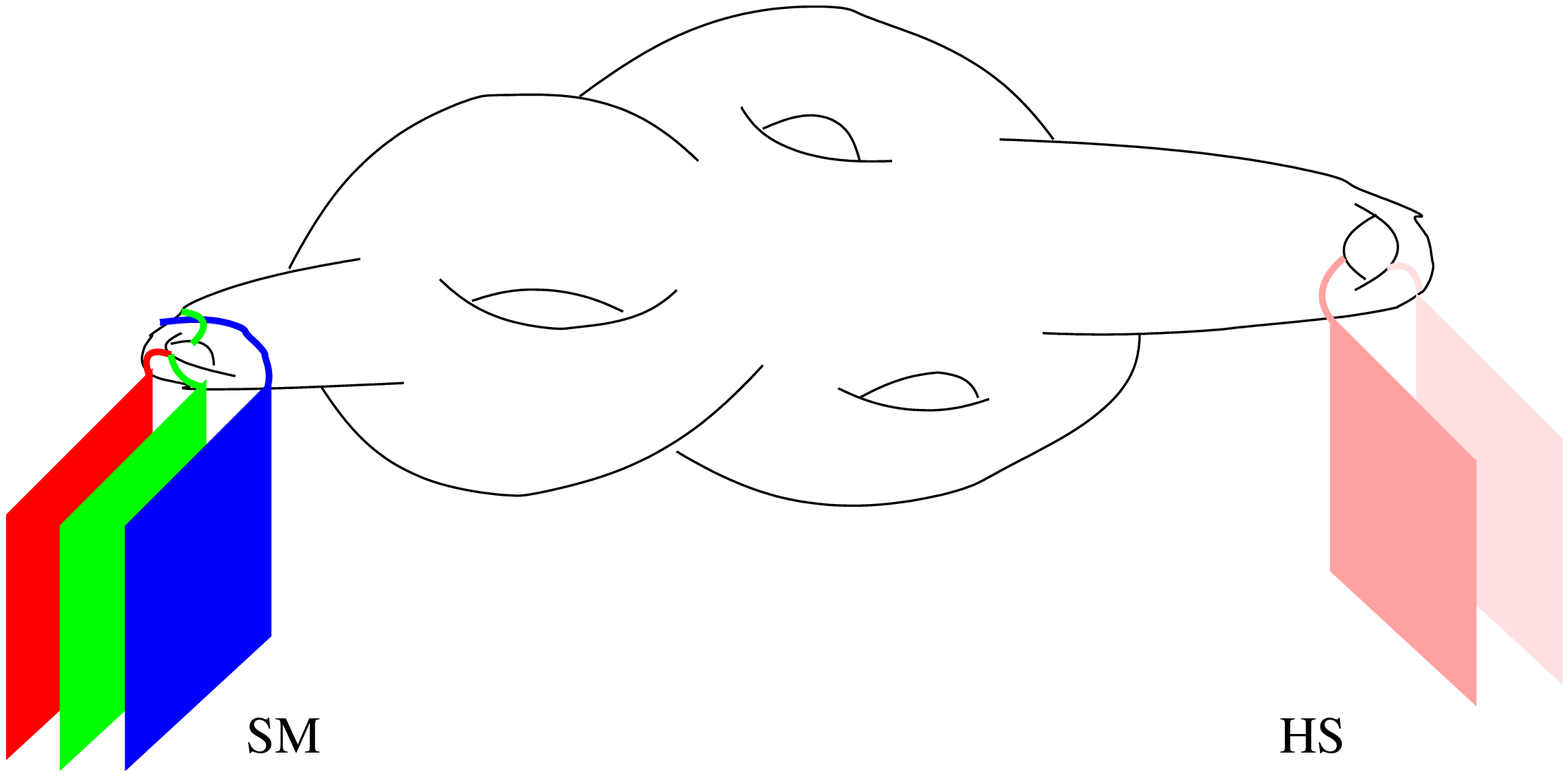}
   }
\end{picture}
\end{center}

\vskip3.6cm

Many details on the construction of open string intersecting brane models and how to relate
the to the SM can be found in various review articles   \cite{Angelantonj:2002ct,Kiritsis:2003mc,Lust:2004ks,Blumenhagen:2005mu,Blumenhagen:2006ci,Marchesano:2007de}.
Let us therefore just summarize the main aspects of the intersecting D6-brane
models.
\begin{itemize}
\item 
We assume that six spatial directions are described
by a compact space  ${\cal M}_{6}$. 
In addition, a consistent orientifold projection is performed.
This yields O6-planes and in general changes the geometry.
The bulk space-time supersymmetry is reduced to ${\cal N}=1$ by the orientifold
projection.
To be more specific we will consider
a type IIA orientifold background of the form
\begin{eqnarray}
{\cal M}^{10} = ({\mathbb R}^{3,1} \times {\cal M}_{6})/
( \Omega\overline\sigma)\, ,\quad \Omega : {\rm world~ sheet ~parity.}
\end{eqnarray}
Here ${\cal M}_{6}$ is a Calabi-Yau 3-fold with a symmetry under 
$\overline\sigma$, the complex conjugation 
\begin{equation}{
\overline\sigma : z_i \mapsto \bar{z}_i,\ i=1,\, ...\, ,3,
}
\end{equation}
in local coordinates $z_i=x^i+iy^i$.
 It is combined with the world sheet parity $\Omega$ 
to form the orientifold projection $\Omega \overline\sigma$. 
This operation is actually a symmetry of the type IIA string 
on ${\cal M}_{6}$. 
Orientifold 6-planes are defined as the 
fixed locus 
\begin{center}
{ ${\mathbb R}^{3,1} \times{\rm Fix}(\overline\sigma)=
{\mathbb R}^{3,1} \times\pi_{O6}
$,} 
\end{center}
where   ${\rm Fix}(\overline\sigma)$ is a supersymmetric (sLag) 3-cycle on
$ {\cal M}^{6}$, denoted by $\pi_{O6}$.
It is special Lagrangian (sLag) and calibrated with 
respect to the real part of the holomorphic 3-form $\Omega_3$.

\vskip0.2cm
\noindent
Next we introduce {D6-branes}  with world-volume
\begin{center}
${\mathbb R}^{3,1} \times\pi_a$, 
\end{center}
i.e. they are
wrapped
around the {supersymmetric (sLag) 3-cycles $\pi_a$ 
and their $\Omega\overline\sigma$ images $\pi_a'$}
of ${\cal M}_6$,
which intersect  in ${\cal M}_6$.
Since the D-branes will
be wrapped around compact cycles
of the internal space,
multiple intersections will now be possible.
The chiral massless spectrum indeed is completely fixed
by the topological { intersection numbers $I$
of the 3-cycles} of the configuration.
\vskip 0.5cm
\begin{tabular}{lll}

\hspace{2cm} 
&
\hspace{2.5cm} 
&
\\

{\bf Sector} & {\bf Rep.} & {\bf Intersection number I} \\[0.2cm]
\hline
\hline
$a'\,a$
&
$A_a$
&
${1\over 2}\left(\pi'_a\circ\pi_a+\pi_{O6}\circ\pi_a\right)$
\\[0.2cm]

$a'\, a$
&
$S_a$
&
${1\over 2}\left(\pi'_a\circ\pi_a-\pi_{O6}\circ\pi_a\right)$
\\[0.2cm]

$a\,b$
&
$(\overline{N}_a,N_b)$
&
$\pi_{a}\circ \pi_b$
\\[0.2cm]

$a'\, b$
&
$({N}_a,{N}_b)$
&
$\pi'_{a}\circ \pi_b$
\\[0.2cm]
\hline
\end{tabular}
\vskip 0.4cm

\item Since
the Ramond charges
of the space-time filling D-branes cannot `escape'
to infinite, the internal Ramond charges on compact space must cancel
(Gauss law). This is the issue of Ramond tadpole cancellation
which give some strong restrictions on the allowed D-brane
configurations. Specifically, the Ramond tadpole conditions follow from the equations
of motion for the gauge field $C_7$:
\begin{equation}
{ {1\over \kappa^2}\,
d\star d C_{7}=\mu_6\sum_a N_a\, \delta(\pi_a)+
                    \mu_6\sum_a N_a\, \delta(\pi'_a)
        + \mu_6 Q_6\,  \delta(\pi_{{\rm O}6}),
}
\end{equation}
where $\delta(\pi_a)$ denotes the Poincar\'e dual form of $\pi_a$, 
$\mu_p = 2\pi (4\pi^2\alpha')^{-(p+1)/2}$, and 
$2 \kappa^2= \mu_7^{-1}$.
Upon
integrating over ${\cal M}_6$ one obtains the RR-tadpole cancellation 
as equation
in homology:
\begin{equation}
\label{tadpoles}
\sum_a  N_a\, (\pi_a + \pi'_a)-4\pi_{{\rm O}6}=0.
\end{equation}
In principle it involves as many linear relations as there are
independent generators in $H_3({\cal M}_{6},R)$. But, of course,
the action of $\overline\sigma$ on ${\cal M}_{6}$ also induces an action
$[\overline\sigma]$ on the homology and cohomology. In particular,
$[\overline\sigma]$ swaps $H^{2,1}$ and $H^{1,2}$, and 
the number of conditions is halved.

\item Next, there
is the requirement of cancellation 
of the internal D-brane tensions, i.e
the forces between the D-branes must be balanced.
In terms of string amplitudes, it means that all NS tadpoles must
vanish, namely all NS tadpoles of the closed string moduli
fields and also of the dilaton field. Absence of these tadpoles means
that the potential of those fields is minimized. 
The disc level tension can be determined by 
integrating the Dirac-Born-Infeld effective
action. It is proportional to the
volume of the D-branes and the O-plane, so that the
disc level scalar potential reads
\begin{eqnarray}
\label{sp}
{\cal V}&=&T_6\, {e^{-\phi_{4}}
\over \sqrt{{\rm Vol({\cal M}_{6})}}}
               \left( \sum_a  N_a \left( {\rm Vol}({\rm D}6_a) + 
      {\rm Vol}({\rm D}6'_a) \right) -4 {\rm Vol}({\rm O}6)\right)\nonumber \\
     &=&T_6\, e^{-\phi_{4}} \left(
\sum_a{N_a \left| \int_{\pi_a} \widehat\Omega_3 \right|} +
 \sum_a{N_a \left| \int_{\pi'_a} \widehat\Omega_3 \right|} 
-4 \left| \int_{\pi_{{\rm O}6}} \widehat\Omega_3 \right |\right) .
\end{eqnarray}
The potential is easily seen to be positive semidefinite and its 
minimization imposes conditions on some of the moduli, freezing them 
to fixed values. Whenever the potential is non-vanishing, supersymmetry 
is broken and a classical vacuum energy generated by the net brane tension. 
It is easily demonstrated that the vanishing of ${\cal V}$ requires all 
the cycles wrapped by the D6-branes to be calibrated with respect to the same 
3-form as are the O6-planes.

\item
One can show that
the cancellation of
the RR tadpoles implies absence of 
the non-Abelian anomalies in the effective 4D field theory.
However there can be still anomalous $U(1)$ gauge symmetries
in the effective 4D field theory.
These anomalies will be canceled by a Green-Schwarz mechanism
involving Ramond (pseudo)scalar field. As a result of these interactions
the corresponding $U(1)$ gauge boson will become massive.
Considering the relevant triangle diagrams the condition for an anomaly free $U(1)_a$ is:
\begin{eqnarray}
N_a(\pi_a-\pi_a')\circ \pi_b=0\, .
\end{eqnarray}
Note that 
even an anomaly free $U(1)$ can become massive.
The massive $U(1)$ always remains as a global symmetry.
For SM engineering, we always have to require that the linear combination of $U(1)$'s that
corresponds to $U(1)_Y$ is anomaly free and massless.

\item
Besides the local triangle anomalies, field theoretical models can be plagued by global
$SU(2)$ gauge anomalies. In orientifold models this requirement can be deduced from a K-theory
analysis. In the case of our models, this condition requires an even amount 
of chiral matter  from $Sp(2)$ probe branes. In this case we obtain the following condition
for a model with $k$ stack of branes:
\begin{equation}
\sum_{a=1}^kN_a\pi_a\circ\pi_p\equiv 0\quad{\rm mod}\quad 2\, .
\end{equation}
This equation should hold for any probe brane $p$ invariant under the orientifold map.

\end{itemize}

To be specific, we now restrict ourselves on orbifold compactifications, i.e.
 ${\cal M}_6$ is a toroidal $Z_N$ resp. $Z_N\times Z_M$ orientifold. First, we
 consider the case  
${\cal M}_6=T^6/Z_2\times Z_2=\prod_{I=1}^3T^2_I/Z_2\times Z_2$. The D6-branes
are wrapping special Langrangian 3-cycles, which are products of 1-cycles
in each of the three subtori $T^2_I$. Hence they are characterized 
by three pairs of integer-valued wrapping numbers $X^I,Y^I$ ($I=0,\dots ,3)$.
The supersymmetry conditions, being equivalent to the vanishing of
the D-term scalar potential ${\cal V}$ (see eq.(\ref{sp})), now have the form:
\begin{equation}\label{dterms}
\sum_{I=0}^3{Y^I\over U_I}=0\, ,\quad\sum_{I=0}^3X^IU_I>0\, .
\end{equation}
The $U_I$ are the three complex structure moduli of the three two-tori $T^2_I$.
The Ramond tadpole cancellation conditions eq.(\ref{tadpoles}) for $k$ stacks
of $N_a$ D6-branes are now given by
\begin{equation}\label{rtadpole}
\sum_{a=1}^kN_a\vec X_a=\vec L\, ,
\end{equation}
where the $L^I$ parametrize the orientifold charge. In addition there are 
some more constraints from K-theory.
Chiral matter in bifundamental representations originate from
open strings located at the intersection of two stacks of D6-branes
with a multiplicity  (generation) number
given by the intersection number
\begin{equation}\label{intersection}
I_{ab}=\sum_{I=0}^3(X^I_aY^I_b-X^I_bY^I_a)\, .
\end{equation}

Now we come to the statistical properties of this particular class of orientifold compactifications.
Specifically, we first want to count all different, consistent D-brane embeddings
into the given $T^6/Z_2\times Z_2$ background geometry. I.e. we want to
count all possible solutions of the D-brane equations (\ref{dterms}) and (\ref{rtadpole}).
These set of equations are diophantic equations in the integer wrapping numbers $X^I,Y^I$,
and they contain as continuous parameters the complex structure moduli $U_I$.
First we want to know, if for any given tadpole charge
$\vec L$ there is a finite number of solutions of these equations.
Actually, based on a saddle point approximation, the total number of D-brane embeddings
can be estimated as follows \cite{Blumenhagen:2004xx}:
\begin{equation}
N_{\rm D-branes}(L)\simeq e^{2\sqrt{L\log L}}\, .
\end{equation}
For typical orientifold charges like $L=64$, one  obtains as estimate
that $N_{\rm D-branes}\simeq 2\times 10^{9}$.

Next, we explicitly count all possible solutions of the D-brane
equations (\ref{dterms}) and (\ref{rtadpole}) by running a computer program; this leads to a total of
$1.66\cdot 10^{8}$ supersymmetric  D-brane models
on the $Z_2\times Z_2$ orientifold \cite{Gmeiner:2005vz,Gmeiner:2005nh,Gmeiner:2006qw}. However  this computer count was limited
by the available CPU time of about $4\times 10^5$ hours, and hence it could be done
only for restricted, not too large values of the complex structure parameters $U_I$.
However, in \cite{Douglas:2006xy} an analytic proof was found that the number of solutions
for eqs.(\ref{dterms}) and (\ref{rtadpole}) is  indeed finite.

With this large sample of models we can ask the question which fraction of 
models satisfy several phenomenological constraints that gradually
approach the spectrum of the supersymmetric MSSM. This is summarized
in the following table:

\vskip4cm

\begin{table}[htb!]
\begin{center}
\begin{tabular}{|l|r|}\hline
Restriction                    & Factor\\\hline
gauge factor $U(3)$            & $0.0816$\\
gauge factor $U(2)/Sp(2)$      & $0.992$\\
No symmetric representations   & $0.839$\\
Massless $U(1)_Y$	       & $0.423$\\
Three generations of quarks   ($I_{ab}^{\rm quarks}=3$) & $2.92\times10^{-5}$\\
Three generations of leptons  ($I_{ab}^{\rm leptons}=3$) & $1.62\times10^{-3}$\\\hline
\emph{Total}                   & $1.3\times10^{-9}$\\\hline
\end{tabular}
\end{center}
\end{table}

The total probability of $1.3\times 10^{-9}$ is simply obtained multiplying
each probability factors in the first six rows, since one can show that there
is little correlation between these individual probabilties.
We see that statistically  only one in a billion 
models give rise to an MSSM like D-brane vacuum.
Multiplying this result with the initial number of models, the chance to find the MSSM is
less then one. One can now compare this statistical result with the explicitly constructed
intersecting D6-brane models with MSSM like spectra (see \cite{Blumenhagen:2006ci} for references
on various intersecting D-brane models). In fact, in \cite{Marchesano:2004xz} a 
$Z_2\times Z_2$ orientifold model with MSSM like spectrum was found that should be
contained in the statistical search discussed above. However unfortunately this model is
outside the range of complex structure moduli  covered by our computer scan.
Also note that all $Z_2\times Z_2$ MSSM like models found so far contain also exotic particles, not
present in the MSSM. These exotic particles were  allowed in our statistical search.

The statistical scan was extended in \cite{Gmeiner:2007we} to the case of the $Z_6$ orbifold geometry
background geometry. For this class of orbifold backgrounds explicit MSSM
like models were constructed in \cite{Honecker:2004kb}.
Compared to the $Z_2\times Z_2$ orientifold, the $Z_6$ case is more complex, because
it also contains exceptional, twisted (blowing-up) 3-cycles, besides the untwisted bulk 3-cycles.
D6-branes wrapped around the exceptional 3-cycles correspond to fractional branes.
First, it was possible to show that even in the presence of the exceptional cycles the
number of the D-brane solutions of the tadpole plus supersymmetry conditions is finite.
Then, by extended computer scan it was found that there exist $3.4\times 10^{28}$ solutions
in total, of which $5.7\times 10^6$ contain the gauge group and the chiral
matter content of the MSSM. We therefore obtained a probability of $1.7\times 10^{-22}$ to
find MSSM like vacua, a number considerably lower than the value $10^{-9}$ for
the case of the  $Z_2\times Z_2$ orientifolds.

Finally, similar results can be obtained for orbifold models with $SU(5)$ GUT gauge
group \cite{Gmeiner:2006vb}.

Complementary to the orbifold models, there are the intersecting brane constructions based
on rational conformal field theories, in particular Gepner models \cite{Blumenhagen:2004cg,Dijkstra:2004ym}
In  \cite{Dijkstra:2004cc,Anastasopoulos:2006da} an extended investigation of the
statistics of Gepner models orientifolds was presented. Here the likelihood
to find MSSM like solutions based on the 4-stack quiver shown before within all
brane solutions is about $10^{-12}$. Note however that in this Gepner model scan
exotic representations were excluded. Recently, 
the search for D-instanton generated righthanded Majorana neutrino masses was included
into the Gepner model statistics \cite{Ibanez:2007rs}.

Finally let us mention that other statistical aspects of the D-brane landscape were discussed in
\cite{Dienes:2006ca}.

\section{Moduli stabilization}\label{flux}

In this section we  discuss a few aspects about the moduli 
stabilization process due to background fluxes  (see ref. \cite{Grana:2006is,Denef:2007pq} for reviews on flux compactifications) and
non-perturbative effects.
The number of flux vacua
on a given CY background space is very huge \cite{Bousso:2000xa,Douglas:2003um,Ashok:2003gk}:
${\cal N}_{vac}\sim 10^{500}$.
Again one can try to make some interesting statistical predictions within the
flux landscape, like the question what is the likelihood for obtaining a tiny cosmological
constant, or if supersymmetry is broken at high or low energy scales \cite{Dine:2004is,Denef:2004cf,Arkani-Hamed:2004fb,Arkani-Hamed:2005yv}.
Of course, in order to make more concrete predictions in the string landscape, the flux vacua statistics
must be eventually combined  with the D-brane
statistics, described before.

To be specific we discuss type IIB  compactifications, which include D3/D7-branes and
the associated O3/O7-planes (class (ii) brane models).
First consider
Ramond and NS 3-form fluxes through 3-cycles
of a Calabi-Yau space ${\cal M}_6$. They give rise to the following effective
flux superpotential in four dimensions 
\cite{Gukov:1999ya,Taylor:1999ii,Giddings:2001yu}:
\begin{equation}
W_{\rm flux}(\tau,U)
\sim \int_{{\cal M}_6}(H_3^R+\tau H_3^{NS})\wedge\Omega\, .
\end{equation}
$W_{\rm flux}$ depends on the dilaton $\tau$ and also on the complex structure moduli
$U$. However, since $W_{\rm flux}$ does not depend
on the K\"ahler moduli, one needs additional
non-perturbative 
contributions to the superpotential in order to stabilize all moduli, as it was proposed
in the KKLT \cite{Kachru:2003aw} scenario. In the following we want to
address the question, what are the restrictions to the KKLT scenario in order to
be implemented in  CY compactifications, and on which concrete CY spaces these 
constraints can be satisfied, such that all moduli
can be indeed fixed.

The non-perturbative part of the KKLT superpotential  is provided by
Euclidean D3-instantons \cite{Witten:1996bn}, 
which are wrapped around 4-cycles (divisors)
$D$ inside ${\cal M}_6$,
and/or gaugino condensations in hidden gauge group sectors
on the world volumes of D7-branes,
which are also wrapped around certain
divisors $D$. 
Both give rise to terms in the superpotential of the form
\begin{equation}
W_{\rm n.p.}\sim g_i\Phi^ne^{-a_iV_i}\, ,
\end{equation}
where $V_i$ is the volume of the divisor $D_i$, depending on the
K\"ahler moduli $T$. 
The fields $\Phi$ are matter fields in bifundamental representations that are located at the intersections of
space-time filling D7-branes, which are at the same time also
intersected by the D3-instantons, resp. the D3-instantons lie
on top of the D7-branes.  Their presence is in general required in order
to make the non-perturbative superpotential invariant under global $U(1)$ symmetries, which are 
remnants of local $U(1)$ gauge symmetries that became massive by the GS-mechanism.
In fact, the $U(1)$ invariance of $W_{\rm n.p}$ can be verified by knowing the transformation properties
of the moduli $V_i$ (shift symmetries) and the matter fields $\Phi$ under the $U(1)$
transformations 
\cite{Achucarro:2006zf,Blumenhagen:2006xt,Haack:2006cy}.
For a pure D3-instanton, which is not intersected by any D7-brane, 
there are no matter fields in $W_{\rm n.p.}$, i.e. $n=0$.

For gaugino condensation in a hidden gauge group, $V_i$ is the 
(holomorphic) gauge coupling constant of $G_{\rm hidden}$, and
$W_{\rm n.p.}$ corresponds to the field theory Affleck-Dine-Seiberg/Taylor-Veneziano-Yankielowicz (ADS/TVY) superpotential \cite{Affleck:1983rr,Taylor:1982bp}.
Here the number of matter fields in  $W_{\rm n.p.}$ is determined by the number of
colors and flavors of $G_{\rm hidden}$. In the simplest case with $G_{\rm hidden}=SU(N_C)$
and $N_F=N_C-1$, $W_{\rm n.p.}$ is induced by a single D3-instanton that is wrapping
the same 4-cycle as the $N_c$ gauge D7-branes.

Note that the pre-factor $g_i$ is in general  not a constant,
but rather depends on the complex structure moduli $U$ \cite{Ganor:1998ai}.
In fact, in case of gauge instantons resp. gaugino condensation $g_i$ is related
to the holomorphic part of the one-loop gauge threshold corrections $\Delta_{\rm 1-loop}(U)$ in 
D-brane models \cite{Lust:2003ky, Akerblom:2007np}:
\begin{equation}
g_i\simeq\exp(\Delta(U))\, .
\end{equation}
Explicit calculations in type IIA intersecting D6-brane models give for ${\cal N}=2$
supersymmetric brane sectors that (here in type IIB notation)
\begin{equation}
\Delta(U)\simeq\log\eta(U)\, ,
\end{equation}
whereas in ${\cal N}=1$ brane sectors the holomorphic threshold corrections
is just a constant. However note that the full, 1-loop threshold contain 
additional non-holomorphic, moduli dependent contributions. They are due to $\sigma$-model
anomalies, and are related to the tree level K\"ahler potential of the matter fields
\cite{Lust:2004cx,Akerblom:2007uc}.

The generation of a non-perturbative
superpotential crucially depends on the D-brane zero modes
of the wrapping divisors, i.e. on the topology of the
divisors together with their interplay with the O-planes and
also with the background fluxes   
\cite{Gorlich:2004qm,Tripathy:2005hv,Bergshoeff:2005yp,Lust:2005cu}.
In case there are too many zero modes (too much supersymmetry is preserved
on the divisor) the non-perturbative superpotential is simply absent.
In gauge theory language, additional (bosonic) zero modes correspond to massless adjoint matter fields, which make gaugino condensation impossible,

It is worth mentioning that recently it became possible to compute the D-instanton generated
superpotential by open string CFT methods \cite{Blumenhagen:2006xt,Ibanez:2006da,Florea:2006si,Akerblom:2006hx,Bianchi:2007fx,Cvetic:2007ku,Argurio:2007vq,Bianchi:2007wy,Antusch:2007jd,bclrw},
namely by computing open string amplitudes between
zero mode fields, which correspond to open strings with ends on the internal Euclidean D-branes, and physical matter
fields, corresponding to open strings with both ends on space-time filling D-branes.
These CFT computations can be applied  for deriving the non-perturbative ADS/TVY superpotential
from D-instantons (for the case $N_F=N_C-1$ in $SU(N_C)$ gauge theories with
$N_F$ vector-like fundamental matter fields, and also
for $Sp(N_C)$ with $N_F=N_C$ fundamental matter fields, and
for $SO(N_C)$ with $N_F=N_C-3$), and hence are relevant for
moduli stabilization. In addition,
the D-instantons can also generate new matter couplings, like righthanded Majorana neutrino masses, Yukawa couplings, FI-terms or non-perturbative contributions to the gauge kinetic functions,
which are classically forbidden by the massive $U(1)$ symmetries mentioned above.

Let us come back to the problem of moduli stabilization.
In the low energy effective supergravity theory, the moduli are stabilized to discrete values by solving
the 
${\cal N}=1$
supersymmetry conditions
\begin{equation}\label{susy}
D_AW=0\quad {\rm (vanishing~~ F-term)}.
\end{equation}
Then typically the superpotentials of the form $W_{\rm flux}+W_{\rm n.p.}$
lead to stable supersymmetric $AdS_4$ minima. Additional restrictions
on the form of the superpotential 
arise   \cite{Choi:2004sx,Lust:2005dy}
requiring that the mass matrix of all the fields $(S,T,U)$ is
already positive definite in the AdS vacuum (absence
of tachyons), as it is necessary,
if one wants to uplift the AdS vacua to a dS vacuum by a shift
in the potential. 
These conditions cannot be satisfied in orientifold
models without any complex structure moduli, i.e. for Calabi-Yau
spaces with Hodge number $h^{2,1}=0$.
Alternatively one can also look 
for
supersymmetric 4D Minkowski minima which solve eq.(\ref{susy})
\cite{Blanco-Pillado:2005fn,Krefl:2006vu}. 
They may exist
if $W_{\rm n.p.}$ is of the racetrack form. In this case the requirement
that all flat directions are lifted in the Minkowski
vacuum leads to similar constraints as the absence of tachyon 
condition in the AdS case.

In more concrete terms, the moduli stabilization procedure to $AdS_4$ vacua was studied in 
\cite{Denef:2005mm}
for the $T^6/Z_2\times Z_2$ orientifold, with the result that all moduli
indeed can be fixed. Moreover in 
\cite{Lust:2005dy,Reffert:2005mn,Lust:2006zg,Lust:2006zh,Reffert:2006du, Reffert:2007im}
all other $Z_N$ and $Z_N\times Z_M$
orientifolds were studied in great detail, where it turns out  that
in order to have divisors, which contribute to the non-perturbative
superpotential, one has to consider the blown-up orbifold
geometries. Then the divisors originating from the
blowing-ups give rise to D3-instantons and/or gaugino condensates,
being rigid and hence satisfying the necessary topological conditions.  
As a result of this investigation of all possible orbifold models,
it turns out that the $Z_2\times Z_2$, $Z_2\times Z_4$, $Z_4$,
$Z_{6-II}$ orientifolds
are good candidates where all moduli can be completely stabilized to a supersymmetric ADS
vacuum with positive definite scalar mass matrix.

The up-lift to a vacuum with positive cosmological constant (dS-vacuum) and the breaking of space-time
supersymmetry can be  done in several ways. First, the original KKLT scenario \cite{Kachru:2003aw}
proposes to introduce anti-D3-branes, which explicitly break supersymmerty. A second
proposal deals with D7-branes that contain additional supersymmetry breaking F-flux on their
world volumes. In field theory language this corresponds to up-lifting and supersymmetry breaking
by a D-term scalar potential \cite{Burgess:2003ic,Jockers:2005zy,Villadoro:2005yq,Dudas:2006vc,Achucarro:2006zf,Haack:2006cy}.
Since in a non-supersymmetric supergravity groundstate, F- and D-terms
are proportional to each other, the non-vanishing D-term potential implies also the existence
of a non-vanishing F-term, i.e. the original F-term SUSY condition eq.(\ref{susy}) of the ADS-vacuum
has to be relaxed.
Another interesting issue with D-term uplifting is that $W_{\rm n.p.}$ and the D-term potential
must be simultaneously invariant under the massive $U(1)$ gauge symmetries, which is
indeed the case after including the matter fields in $W_{\rm n.p.}$. 
Alternatively, up-lifting by 
F-terms was discussed in the literature  \cite{Dudas:2006gr,Lebedev:2006qc}, where again matter fields play an important rule.
Moreover, the existence of recently discovered metastable, non-supersymmetric vacua  in supersymmetric
gauge theories \cite{Intriligator:2006dd} and in their string realizations \cite{Franco:2006es,Ooguri:2006bg}
most likely will become important for the up-lifting procedure.
However many details of the up-lifting procedure still have to worked out in concrete models.

After breaking space-time supersymmetry the soft SUSY breaking parameters for the matter fields
on the D-branes can be computed \cite{Camara:2003ku,Grana:2003ek,Lust:2004fi,Camara:2004jj,Lust:2004dn,Font:2004cx,Lust:2005bd,Choi:2005ge,Dudas:2005vv,Conlon:2006wz,Conlon:2007xv}. It will become important to compare the pattern of
soft parameters obtained in D-brane models with the experimental data, which we will
hopefully obtain from the LHC in the next years. E.g. the analysis in \cite{Lust:2004dn} indicates
that the squark masses are heavier that the gaugino masses on the D7-branes.

\section{Black hole/flux correspondence and probability wave function}\label{bh}

In this section we like to discuss a method to assign a probability
measure to flux compactifications via an associated black hole entropy functional
${\cal S}$. 
As mentioned in the introduction, it is possible to compute the thermodynamic
Bekenstein-Hawking entropy of charged, supersymmetric black holes in
string theory by counting the associated microscopic string degrees of freedom
(see ref. \cite{Mohaupt:2007mb} for a black hole review).
To be specific, we like consider charged, supersymmetric black holes in 
${\cal N}=2$ supergravity, which arise by compactifying type II A/B superstrings on a
Calabi-Yau manifold. The four-dimensional black holes carry magnetic/electric $U(1)$ charges, denoted by $p^I$ and $q_I$, with respect to the 4D Abelian field strength  $F_I$ and $G^I$ ($I=0,\dots N$):
\begin{equation}
q_I=\oint_{S^2}F_I\, ,\quad p^I=\oint_{S^2}G^I\, 
\end{equation}
where the 2-sphere $S^2$ is embedded into the four-dimensional space time.
In ${\cal N}=2$ supergravity, these $N+1$ fields strengths fields are components
of $N$ vector multiplets plus the graviphoton field strength.
In type IIB compactifications, the ${\cal N}=2$ vector multiplets correspond to the
complex structure deformations, i.e. $N=h^{2,1}$, whereas in type IIA they are
related to the K\"ahler deformations, and therefore $N=h^{1,1}$.
In the following we concentrate on the type IIB case. 
 Here, the four-dimensional field
strength fields arise by dimensionally reducing the type IIB Ramond 5-form field strength $F_5$ on
internal internal three cycles $\Sigma_I\subset{\cal M}_6$ and $\tilde\Sigma^I\subset {\cal M}_6$:
\begin{equation}
F_I=\oint_{\Sigma_I}F_5\, ,\quad G^I=\oint_{\Sigma^I}F_5\, 
\end{equation}
The charged black holes have the interpretation of D3-brane, which are
the sources of the 5-form field strengths $F_5$, and which are  wrapped around
these 3-cycles.

In four dimensional supergravity, the supersymmetric, charged blackholes are interpolating solutions that interpolate
between flat space at radial infinity and the $AdS_2\times S^2$ geometry at the horizon
of the black hole. At the horizon, full ${\cal N}=2$ space-time supersymmetry is restored. Furthermore,
according to the attractor mechanism, the values of the scalar fields $\phi_A$ (complex structure moduli) at the horizon are 
determined by the supersymmetry condition (stabilization condition)
\begin{equation}
D_AZ(\phi)=0\quad\Longrightarrow\quad\phi_{A, {\rm hor.}}\, ,
\end{equation}
where $Z$ is the central charge of the ${\cal N}=2$ supersymmetry algebra.
Solving this condition, it follows that the $\phi_{A, {\rm hor.}}$
are entirely given in terms of the magnetic/electric
charges $(p^I,q_I)$: 
\begin{equation}\label{stab}
\phi_{A, {\rm hor.}}=\phi_A(p^I,q_I)\, ,\quad A=1,\dots , N\, .
\end{equation}
Finally, the Bekenstein-Hawking entropy of the charged black holes is
also a function of the charges $(p^I,q_I)$, as it is given in terms
of the value of the central charge $Z$ at the horizon:
\begin{equation}
{\cal S}_{bh}(\phi(p,q))=|Z(\phi(p,q))|^2={A_{\rm hor}(\phi(p,q))\over 4}\, .
\end{equation}
Using the attractor relation between charges $(p^I,q_I)$ and scalar moduli fields, eq.(\ref{stab}),
the black hole entropy can be also regarded as a functional over
the complex structure moduli space:
\begin{equation}\label{entrop}
{\cal S}_{bh}={\cal S}(\phi)\, .
\end{equation}
However it is important to note that the attractor relation eq.(\ref{stab}) is not unique, i.e. one-to-one,
since there is one more pair of charges $(p^I,q_I)$ compared the number of scalar fields $\phi_A$.
Hence to compute the entropy functional ${\cal S}(\phi)$ from the attractor equations, one
electric and one magnetic charge has to be fixed by hand.

The black hole  attractor equations are quite similar to the supersymmetry conditions
eq.(\ref{susy}) in the context of flux compactifications
\cite{Behrndt:2001mx,Kallosh:2005bj}.
Therefore one might conjecture a
close connection between the string flux landscape and the entropy of charged black holes.
In fact, instead of 
3-form flux vacua in four dimensions, Ooguri, Verlinde, Vafa (OVV) \cite{Ooguri:2005vr,Gukov:2005bg}
considered
Ramond 5-form flux compactifications on $S^2\times {\cal M}_6$ to two dimensions.
The corresponding 2-dimensional, supersymmetric vacua, which follow from a supersymmetry condition
\begin{equation}
DW=0\, ,\quad W\sim\int_{S^2\times {\cal M}_6}(F_5\wedge\Omega)\, ,
\end{equation}
have $AdS_2$ geometry.
This supersymmetry condition is formally the same as  the stabilization condition
$DZ=0$ of the 4D charged black holes discussed above.
In view of this connection, it was suggested by OVV 
that the entropy functional of the 4D black holes determined the  wave function 
of the 2D flux vacua, namely
to interpret
\begin{equation}
\psi(\phi)=e^{{\cal S}(\phi)}
\end{equation}
as a probability distribution resp. wave function for 2D flux
compactifications. So the probability functions is peaked at points of
maximal entropy in the moduli space. In a stringy picture, 
$\psi$ essentially counts the microscopic string degrees
of freedom, which are associated to each flux vacuum.

The correspondence between flux vacua and charged black holes can be 
summarized by the following table:

\vskip0.5cm
\begin{tabular}{c|c}
Black holes    &Flux compactifications\\ \hline
D3-branes wrapped around $\Sigma_3$ &$F_5$ through $S^2\times \Sigma_3$\\
Black hole charges $(q_I,p^I)$ & 5-form fluxes $(q_I,p^I)$\\
Central charge $Z(\phi)$ & Superpotential $W(\phi)$ \\
Stabilization cond.  $D_AZ=0$     &     Supersymmetry cond. $D_AW=0$\\
Entropy ${\cal S}$ & Cosmological constant $-\Lambda$\\
Near horizon geometry $AdS_2\times S^2$ & Vacuum space $AdS_2\times S^2$\\
\end{tabular}  

\vskip0.6cm
As can be seen from this table, the black hole entropy corresponds to the negative
value of the $AdS_2$ cosmological constant in the flux landscape.
Hence one obtains for the wave function $\psi$ the following relation:
\begin{equation}
\psi(\phi)=e^{-\Lambda(\phi)}\, .
\end{equation}
This form of the wave function is very similar to the Hartle-Hawking wave function,
which describes certain tunneling amplitudes between different vacua.
It implies that $AdS_2$ vacua with minimal cosmological constant 
possess the largest probability.

The principle of entropy maximization was further studied in \cite{Gukov:2005bg,Fiol:2006dv,Cardoso:2006nt}.
As a simple toy example the entropy functional near the special
point in Calabi-Yau compactifications, where new states become massless (conifold
point), was studied. 
In particular, the following question was addressed, namely
is the entropy functional maximal resp. minimal in case ${\cal N}=2$ vector/hyper multiplets
become massless at this special point?  For massless hypermultiplets,
the effective gauge theory at the conifold is infrared free (negative $\beta$-function
coefficient, i.e. $\beta<0$, and for massless vector multiplets
the effective gauge theory is asymptotically free, i.e. $\beta>0$.

The answer to the question above is the following.
Using the known expression for the ${\cal N}=2$ prepotential near the conifold point,
the entropy functional takes the following form
\begin{equation}
{\cal S}(\phi)={\rm const.}-{\beta\over 2\pi}({\rm Re}~\phi)^2-{2\beta\over\pi}|\phi |^2\log |\phi |\,.
\end{equation}
Here $\phi$ is the complex structure modulus, which describes the moduli space in
the neighborhood of the conifold point. From this formula it follows that
the entropy is maximal at $\phi=0$ for $\beta<0$ and minimal for $\beta>0$ \cite{Cardoso:2006nt} .
Hence, according to this discussion the effective, infrared free gauge theory is more
probable than the asymptotically free gauge theory.
This conclusion also persists, if one includes higher curvature corrections in the entropy functional,
or if one uses the non-perturbative prepotential from topological string theory 
\cite{Gopakumar:1998ii}.

In \cite{Cardoso:2006cb} the entropy functional of 
non-supersymmetric, charged black holes around the 
conifold point was studied. In turns out that in this case the
conifold point is not anymore an attractor point. However one can show that there exist an
attractor point in the close neighborhood of the conifold point.

Finally let us end with some comments on a probability functional of 4-dimensional 
$AdS_4$ flux vacua.
Supersymmetric $AdS_4$ vacua with all moduli fixed can be constructed in type IIA flux compactifications
with 0,2,4,6-form Ramond fluxes  plus possibly
geometrical fluxes
\cite{Derendinger:2004jn,Villadoro:2005cu,DeWolfe:2005uu, Camara:2005dc,Lust:2004ig,Villadoro:2007yq}, 
and also in type IIB on
some non-geometrical background spaces with 3-form fluxes
\cite{Becker:2006ks,Becker:2007dn}. In the next step one replaces
the various fluxes by their corresponding smeared 
brane sources \cite{KLPT}. These branes are wrapping
internal cycles, and they are domain walls in the four-dimensional space time, like
D4-branes wrapped around internal 2-cycles, D8-branes wrapped around the CY-space
or NS5-branes wrapped around internal 3-cycles.
As one can show \cite{KLPT}, these brane configurations  are supersymmetric, create
the same tadpoles as the fluxes, and in 4D
their interpolate between the near horizon geometry $AdS_4$ and flat Minkowski
space-time. So having constructed  these supersymmetric brane configurations that
correspond to $AdS_4$ flux vacua, one might hope to determine thei
entropies of these 4D domain walls and to get in this way a probability functional in the four-dimensional
flux landscape.

\section{Final comments}\label{final}

In this talk we addressed the question how to relate the string theory landscape with
low energy physics, in particular with the MSSM. Phenomenologically interesting heterotic constructions
and D-brane models were discussed.
In particular starting from
a large sample of consistent D-brane embeddings into
a given 6-dimensional internal space, we investigated the statistical likelihood of obtaining  models
which come to the MSSM. 
In additions, some mechanisms for moduli stabilization were investigated, where we
also addressed the question whether these mechanisms can be successfully be realized
in concrete orientifold compactifications. 
One caveat in this discussion is that the
moduli stabilization occurs within the effective supergravity
approach which is limited to large volumes of
the internal spaces. It would be interesting
to learn more about the string landscape taking into
account higher order $\alpha'$ corrections, or investigating the
problem of the string ground state in a background independent
formulation. Finally, some ideas about probability functions
in the string moduli space in relation with black hole entropies were presented. This apparently
works for two-dimensional $AdS_2$ flux vacua, which are related to 4D charged black hole solutions.
Whether a similar constructions also works for 4-dimensional $AdS_4$ flux vacua, which can
be related to 4D domain wall solutions, still has to be seen in the future.
For the string landscape discussion it is also important to learn what kind of
field theory are inside the string landscape and can be consistently coupled to gravity,
and which field theories lie outside the string landscape, the socalled swampland \cite{Ooguri:2006in}.

During the next years, we hope to get a lot of new informations for string model building from
new experiments, most notable the LHC and new astrophysical observations.
It will be very important to concretely work out the chain between the experimental data and
the string theory parameters, which are to large extent given in terms of geometrical
and topological parameters of the internal compact spaces. 
At the moment string theory still does not make too many very explicit predictions about
soft supersymmetry breaking masses etc. However there are important qualitative predictions
from string theory. In particular we are interested in the answer to the following questions:
\begin{itemize}
\item Where is the intrinsic string scale $M_{\rm string}$?
\item What is the size of the extra dimenensions?
\item What is the scale of supersymmetry breaking?
\end{itemize}
To answer these questions
is indeed a great challenge for the next years;
to make progress in our understanding about string theory or more generally about
physics at very short distances needs big combined efforts
of mathematicians, string theorists, particle phenomenologists, astrophysicists and last but not least
of experimentalists.

\section*{Acknowledgments}
It is a great pleasure to thank my collaborators R. Blumenhagen, G. Lopes Cardoso,
G. Curio, F. Gmeiner, V. Grass, M. Haack, G. Honecker, B. K\"ors,
 D. Krefl, J. Perz, S. Reffert, E. Scheidegger,
W. Schulgin, S. Stieberger, A.  Van Proeyen, D. Tsimpis, T. Weigand and M. Zagermann
for fruitful and pleasant collaboration on the material presented in this paper.
This work is supported in part by the European Community's Human
Potential Program under contract MRTN-CT-2004-005104 `Constituents,
fundamental forces and symmetries of the universe' and also by
'The Cluster of Excellence for Fundamental Physics - Origin and
Structure of the Universe' in M\"unchen

\eject

\vfill

\end{document}